\documentclass[runningheads]{llncs}
\usepackage[T1]{fontenc}
\usepackage{graphicx}
\usepackage{hyperref}
\usepackage{url}
\hypersetup{colorlinks=True, urlcolor=blue}

\usepackage{color}

\usepackage{multirow}
\usepackage{multicol}

\begin{document}
\title{Multi-Speaker Multi-Style Speech Synthesis with Timbre and Style Disentanglement\thanks{This work was supported by the National Key R\&D Program of China under Grant No. 2020AAA0108600.}}
\titlerunning{Multi-Speaker Multi-Style Speech Synthesis}
\author{Wei Song \and
Yanghao Yue \and
Ya-jie Zhang \and
Zhengchen Zhang \and
Youzheng Wu \and
Xiaodong He}
\authorrunning{W. Song et al.}
\institute{JD Technology Group\\
\email{\{songwei11,yueyanghao,zhangyajie23,zhangzhengchen1,\\
wuyouzheng1,hexiaodong\}@jd.com}}
\maketitle              % typeset the header of the contribution

\begin{abstract}
Disentanglement of a speaker's timbre and style is very important for style transfer in multi-speaker multi-style text-to-speech (TTS) scenarios. 
% With the disentanglement of timbre and style, TTS system could synthesize expressive speech with any combination of speaker's timbre and style that seen in the training corpus.
With the disentanglement of timbres and styles, TTS systems could synthesize expressive speech for a given speaker with any style which has been seen in the training corpus.
However, there are still some shortcomings with the current research on timbre and style disentanglement. 
The current method either requires single-speaker multi-style recordings, which are difficult and expensive to collect, or uses a complex network and complicated training method, which is difficult to reproduce and control the style transfer behavior.
% A single-speaker with multi-style recordings is essential to learn the style representation in some works, usually, these kinds of corpus are difficult and expensive to collect. 
% Some of the researches use complex neural networks and training methods, which makes it difficult to reproduce the experimental result and control the style transfer behavior. 
% , then causes poor performance.
To improve the disentanglement effectiveness of timbres and styles, and to remove the reliance on single-speaker multi-style corpus, a simple but effective timbre and style disentanglement method is proposed in this paper.
% In this paper we proposed a simple but effective timbre and style disentangle neural network based on FastSpeech2, each speaker's speech data is considered as a separate and isolated style, then single speaker multi style corpus is not needed.
The FastSpeech2 network is employed as the backbone network, with explicit duration, pitch, and energy trajectory to represent the style. 
Each speaker's data is considered as a separate and isolated style, then a speaker embedding and a style embedding are added to the FastSpeech2 network to learn disentangled representations.
% this remove the dependent of multi-style data for a single speaker;
Utterance level pitch and energy normalization are utilized to improve the decoupling effect. 
% This normalization is especially useful for noisy data, because the noise could be considered as a constant energy value and utterance level normalization canceled this constant value out.
% which also helps for noisy data because the noise could be considered as a constant energy value and utterance level normalization canceled this constant value out.
% Utterance level pitch and energy normalization is used to improve the style similarity, which also helps for found data because the noisy in found data could be considered as a constant energy, by utterance level normalization this constant is canceled out.
% The third part (Instance Normalization) could be removed, because it's not that important
% At last, instance normalization is used in the pitch and energy dimension raising layer, which is found helpful for better timbre and style disentanglement.
Experimental results demonstrate that the proposed model could synthesize speech with any style seen during training with high style similarity while maintaining very high speaker similarity.
% while keeping the speaker's timbre unchanged.
% Both subjective and objective experimental results showed the effectiveness of the proposed method.
\end{abstract}
\section{Introduction}

With the development of deep learning technology in the last decade, speech synthesis technology has evolved from traditional statistics-based speech synthesis~\cite{taylor2009text} to end-to-end based~\cite{sotelo2017char2wav,wang2017tacotron,ren2020fastspeech,ping2017deep,tan2021survey} and made great advancements. 
The current speech synthesis technology has been able to synthesize speech with high naturalness and high fidelity, and even some research~\cite{tan2022naturalspeech} has been able to synthesize speech that human beings cannot distinguish between true recordings.

Although the great achievement in speech synthesis, there still exists a large improvement room for expressive speech synthesis, especially for multi-speaker multi-style Text to Speech (TTS) with cross-speaker style transfer~\cite{pan2021cross,xie2021multi,an2022disentangling}. 
Synthesizing speech with a target speaker's timbre and other speaker's style could further increase the application scenarios and expressiveness of the TTS system.
% , which could also be called style transfer.
% multi speaker multi style speech synthesis with cross speaker style transfer still has a lot of room for improvement. Synthesize speech with a target speaker's timbre and other speaker's style could further increase the application scenarios, user experience and expressiveness of TTS systems. 

In order to synthesize more expressive speech, some researches~\cite{li2021controllable,pan2021cross,an2022disentangling} do style decoupling by using a single-speaker multi-style recording data to learn the style representation.
% , or a multi-speaker corpus where at least one speaker's data should consist of multi-style recordings to learn the style representation. 
However, it is difficult and expensive to collect these kinds of data in lots of scenarios.
%However, in lots of scenarios these kind of data is difficult or expensive to collect.
% Some other works try to remove this data restriction by disentangling speaker's timbre and style, but much of them use complex network structure or relay on complex training method, which make it difficult to re-produce the experimental results or infeasible to control the network output in inference procedure.
Some other works~\cite{wang2018style,xie2021multi} try to remove this data restriction by learning a decoupled style representation, but most of them use complex network structure or rely on complicated training method, which makes it difficult to re-produce the experimental results or infeasible to control the network performance during inference.

% Style transfer works~\cite{wang2018style,whitehill2019multi} use a reference encoder to learn implicit style representation, usually the reference encoder is trained jointly with the TTS model. 
A jointly trained reference encoder is used to learn implicit style representation in~\cite{wang2018style,zhang2019learning}. 
After the model is trained, audio with a different style or even from a different speaker could be taken as the reference audio to synthesize speech with the desired style while keeping the timbre unchanged.
% Paper~\cite{zhang2019learning} incorporates Variational Auto-Encoder (VAE)~\cite{kingma2013auto} into GST-Tacotron~\cite{wang2018style} in order to disentangled style attributes of the style embedding. 
However, the reference-based method is unstable which usually generates unexpected style, and it's non-trivial to choose the reference audio.
% the choice of reference audio is tricky.

% Most of the recent researches on style transfer focus on the disentanglement of the speaker's timbre and style.
% Some other works~\cite{pan2021cross,xie2021multi,an2022disentangling} focus on the disentanglement of timbre and style. 
Pan \textit{et al.}~\cite{pan2021cross} use prosody bottle-neck feature to learn a compact style representation, 
% in inference procedure the speaker embedding should be replaced to target speaker embedding after the prosody bottle-neck feature prediction. 
and paper~\cite{an2022disentangling} uses a separate style encoder and speaker encoder to disentangle the speaker's timbre and style, and cycle consistent loss is used to improve the disentanglement effect, a complex neural network and complicated training objectives are required to achieve good performance. 
Both ~\cite{an2022disentangling} and ~\cite{pan2021cross} require single-speaker multi-style corpus to learning the disentanglement, which restricts the flexibility of their proposed models.
% Xie \textit{et al.}~\cite{xie2021multi} proposed a network that no single-speaker multi-style corpus is required, and used a prosody predictor to predict text-based prosody feature, then a multi-scale prosody encoder is used to extract abstract and multi-scale prosody representation.

In this paper, we propose a simple but effective expressive speech synthesis network that disentangles speakers' timbres and styles, which makes it available to do multi-speaker multi-style speech synthesis.
The proposed system does not require a multi-speaker multi-style corpus, each speaker's data is considered as an isolated style. 
The proposed network is similar to work~\cite{xie2021multi}, but we use FastSpeech2~\cite{ren2020fastspeech} as the network backbone, which removes the skip/repeat pronunciation issue caused by the attention mechanism. 
% and the multi-scale prosody encoder is removed, 
Prosody features (duration, pitch, and energy) are used directly to improve the disentanglement effectiveness of timbre and style, furthermore, utterance level pitch and energy normalization (UttNorm) are used to prevent identity leakage from prosody features.

% The speech synthesis with other speaker's style is also called style transfer, in which the target speaker lack the style data and need to learning style from other speaker's training data.

% style fade from source speaker's style and target speaker's style

% \subsection{A Subsection Sample}
% Please note that the first paragraph of a section or subsection is
% not indented. The first paragraph that follows a table, figure,
% equation etc. does not need an indent, either.

% Subsequent paragraphs, however, are indented.

\section{The Proposed Model}

% The proposed network uses FastSpeech2~\cite{ren2020fastspeech} as the network backbone, and an extra style embedding is introduced to control the style features prediction and disentangle style from timbre. 
% To achieve a better decoupling effect, utterance level pitch and energy normalization are proposed. 
The proposed network utilizes FastSpeech2~\cite{ren2020fastspeech} as the network backbone and applies utterance level pitch and energy normalization to achieve a better decoupling effect. 
MelGAN~\cite{kumar2019melgan} is used as the neural vocoder to convert acoustic features to speech.

\subsection{The Network Structure Of Proposed Network}

\begin{figure}[t]
\vspace{0mm}
\centering
\includegraphics[width=\linewidth]{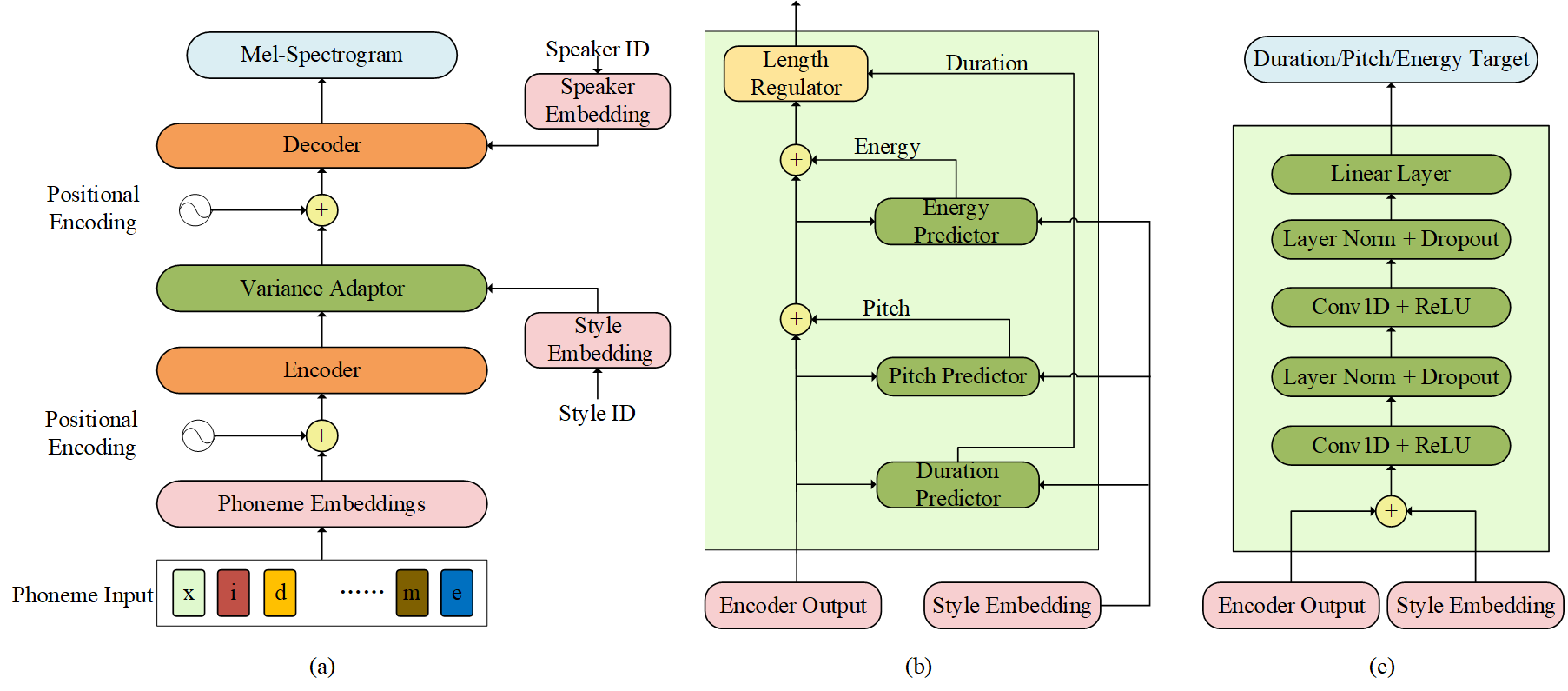}
\caption{An illustration of the proposed network architecture. (a) Network structure of the acoustic model. (b) The variance adaptor structure with style embedding as conditional input. (c) The structure of variance predictor.}
\vspace{-3mm}
\label{network_structure}
\end{figure}

FastSpeech2~\cite{ren2020fastspeech} network architecture is used as the network backbone for the proposed model, which consists of a phoneme encoder to learn syntactic and semantic features, a mel-spectrogram~\cite{wang2017tacotron} decoder to generate frame-level acoustic features, and a variance adaptor to learn style-related features. 
The network structure is shown in Fig.~\ref{network_structure} (a).
%Inspired by DelightfulTTS~\cite{liu2021delightfultts}, to learn both local and global contextual features, we replaced the original feed-forward network in the Transformer block~\cite{vaswani2017attention} to two 1-dimensional convolutional (Conv1d) layers with kernel size 9 and 1 respectively. 
%More details of the network structure could be found in ~\cite{ren2020fastspeech} and ~\cite{liu2021delightfultts}.

% The variance adaptor learns duration, pitch, and energy features to control the speech style, 
A style embedding is introduced to the network to learn a style-dependent variance adaptor, the network structure of variance adapter is illustrated in Fig.~\ref{network_structure} (b) and Fig.~\ref{network_structure} (c). 
%Phoneme-level pitch and energy values are used in this work, and they perform better than frame-level values. 
To decouple timbre and style, the speaker embedding is moved to the input of the decoder to make the decoder timbre dependent. 
%As the style embedding controls the style prediction and makes the variance adaptor style dependent, in order to decouple timbre and style, we further move the speaker embedding to the input of the decoder to make the decoder timbre dependent. 

One important purpose of the proposed model is to remove the dependency of the single-speaker multi-style corpus, so each speaker's corpus is considered as a unique style and we could learn style representation from other speakers' corpus. 
%The style would be more expressive and natural if the speaker records the audio in his talented domain. 
Actually, during the training procedure, the speaker id and style id is identical, then it is very important to prevent style embedding from leaking into the backbone network, which means style embedding should only be used for style feature prediction and never be exposed to the backbone network. 
So in this proposed network, style embedding is only used in the variance adaptor to ensure that the whole network learns the speaker's timbre by speaker embedding instead of style embedding, and style is only affected by style embedding.

% For the style features, duration values, pitch and energy trajectory are scalar values, which prevent information leakage to backbone network. However, these scalar trajectory could still contains speaker identity and impact the decoupling effect, so utterance level normalization is further introduced to improve the decoupling effect, which is explained in detail in next section.

%The duration value, pitch, and energy trajectory are scalar values, which prevent speaker information leakage to the backbone network. 
%For the style features, the duration value, pitch, and energy trajectory, these scalar trajectories could still contain speaker identity and impact the decoupling effect, so utterance level feature normalization is further introduced to improve the decoupling effect, which is explained in detail in the next section.

% \begin{figure}
% \centering
% \includegraphics[width=0.8\textwidth]{figs/variance_adapter.png}
% \caption{The architecture of variance adaptor. (a) The variance adaptor structure with style embedding as conditional input. (b) The structure of variance predictor.}
% \vspace{-2mm}
% \label{variance_adaptor}
% \end{figure}

% FastSpeech2~\cite{ren2020fastspeech} uses pitch and energy quantization and a lookup table to convert a scalar value to embedding, however, in this work we compared quantization with the lookup table method and a learned Conv1d layer for dimension increasing and found that Conv1d layer performs better. 
% For this Conv1d layer, layer normalization~\cite{ba2016layer} and instance normalization~\cite{ulyanov2016instance} are compared for decoupling effect, and instance normalization is found more effective.

\subsection{Utterance Level Feature Normalization}

The speaking style is represented in many aspects, such as the duration of each syllable, the fundamental frequency (F0), and the trajectory of pitch and energy. However, these scalars could still contain speaker identity and impact the decoupling effect. 
%In this paper, phoneme duration, pitch, and energy trajectory are used explicitly to learn a speaker's style. 
For example, female speakers usually have higher F0~\cite{zen2009statistical} than male speakers, and excited speakers express higher energy value, so the pitch and energy features contain speaker timbre information to some extent, then it's essential to normalize the style features to disentangle a speaker's timbre and style.

% In this paper, utterance level pitch and energy normalization (UttNorm) is used to remove speaker identity from these style features for better timbre and style disentanglement. 
% It makes a bigger difference for unprofessional, noisy user recordings which have a significant variation than speaker level normalization.

In this paper, instead of speaker level normalization (SpkNorm), utterance level pitch and energy normalization (UttNorm) are used to remove speaker identity from the style features for better timbre and style disentanglement. 
The style difference in each utterance will lead to a statistics difference between speaker level and utterance level statistics, this difference could cause identity leakage and affect the decoupling effect, especially for the unprofessional, noisy recordings. 
UttNorm could eliminate this statistic difference and improve the timbre and style disentanglement effectiveness.

\section{Experimental Setup}

% \subsection{Training Data Processing}

% To evaluate the performance of the proposed model, 
Three open-sourced Chinese mandarin corpora and three internal Chinese mandarin corpora with distinctive styles are used to train the proposed model.
The open-sourced corpus includes CSMSC~\footnote{\url{https://www.data-baker.com/open_source.html}}, which is recorded by a female speaker, and the MST-Originbeat~\footnote{\url{http://challenge.ai.iqiyi.com/detail?raceId=5fb2688224954e0b48431fe0}}~\cite{xie2021M2VoC}, which consists of recordings from a female speaker and a male speaker. 
% As these three public corpora' style is not significant enough, we added three internal speakers' corpus to show the disentanglement effect, the styles of these three corpora are story-telling, children story, and news-broadcasting style respectively. 
Details of the training data are listed in table~\ref{tab:data_dest}.

\begin{table}[t]
  \caption{The description of the dataset used in the experiment. There are three open-sourced corpus (with a * in the speaker name) and three internal corpora.}
%   \vspace{3pt}
  \label{tab:data_dest}
  \centering
  \begin{tabular}{lrrr}
  \hline
    Speaker         &\#Utterance   & Style                  & Gender       \\
    \hline
    CSMSC*          & 10000        & normal                 & ~~~female    \\
    \hline
    Originbeat-S1*  & 5000         & normal                 & female       \\
    \hline
    Originbeat-S2*  & 5000         & normal                 & male         \\
    \hline
    C1              & 10000        & children story         & female       \\
    \hline
    F1              & 450          & ~~~~news-broadcasting     & female       \\
    \hline
    M1              & 35000        & story-telling          & male         \\
    \hline
  \end{tabular}
  \vspace{-5mm}
\end{table}

The training data waves are converted to 16kHz, 16bit depth, and then scaled to 6dB in our experiments. 
% Redundant silence of each utterance is removed in the beginning and ending part of each file. 
% The text for each wave is converted into phoneme sequence by an internal tool.
The extracted phoneme labels and processed speech waves are aligned by the MFA~\cite{mcauliffe2017montreal} tool to detect the phoneme boundary.
The 80-band mel-scale spectrogram is extracted as the training target with a 12ms hop size and 48ms window size.  
% and the extracted mel-spectrogram is min-max normalized~\cite{lecun2015deep}. 
Pitch is extracted by using the  PyWORLD~\footnote{\url{https://github.com/JeremyCCHsu/Python-Wrapper-for-World-Vocoder}} toolkit. 
Both pitch and energy trajectories are normalized in utterance level to remove speaker identity.

% \subsection{Network Configuration And Training Details}

% A variant version of FastSpeech2~\cite{ren2020fastspeech} network is used in the proposed model, detailed network structure and configuration could be found in paper ~\cite{cui2021emovie}. 
% As the proposed network takes style id as input, two separate lookup tables are used to learn separate style embedding and speaker embedding, the style embedding dimension is identical to speaker embedding.

% Although single-speaker single-style corpus is used in the experiments, it is easy to extend the proposed network to single-speaker multi-style corpus, by simply treating each style of the speaker as a new speaker.

% Each speaker's corpus has a unique speaking style, so in the training procedure, speaker name could also represent the speaker's style, the speaker's name is used twice for two separate embedding lookup tables, named speaker embedding and style embedding respectively. 

The proposed model is trained by Adam~\cite{kingma2014adam} optimizer with a batch size of 32 and noam~\cite{vaswani2017attention} learning rate schedule. The learning rate is warmed up to a maximum value of 1e-3 in the first 4000 steps and then exponentially decayed. The model is trained by 400,000 steps and the network is regularized by weight decay with a weight of 1e-6. 

% The neural vocoder~\cite{kumar2019melgan} is also trained with the same data, a constant learning rate of 1e-4 is used for the first 1M steps training, and then the learning rate is halved and the model is trained by another 1M steps.

\section{Experimental Results}

This chapter shows the experimental results of the proposed timbre and style disentanglement network. %firstly we show the Mean Opinion Score (MOS) results of synthesized speech with different styles and different speakers; 
%and then experimental results with noisy data are shown to verify the effectiveness of utterance level normalization; 
%thirdly, the mel-spectrogram of synthesized speech is illustrated to demonstrate the performance of the proposed multi-speaker and multi-style model; 
%at last, a style transition example is given to show the representation ability of learned style embedding. 
We encourage the readers to listen to the synthesized speeches on our demo page~\footnote{\url{https://weixsong.github.io/demos/MultiSpeakerMultiStyle/}}.

% \subsection{Subjective Evaluation Of Speaker And Style Similarity}

\subsection{Subjective Evaluation}

To evaluate the proposed method, Mean Opinion Score (MOS) evaluation is conducted to evaluate the speaker similarity and style similarity for the synthesized speech of different speakers and different styles.
The target speakers are from the 3 open-sourced corpora, and our internal news-broadcasting (F1), children story (C1), and story-telling (M1) styles are used as the target style. 
Twenty utterances are synthesized for each speaker and style combination, listened to by 15 testers. 
%A 5 points MOS evaluation is used for this subjective evaluation, where 5 means the style or the timbre is exactly the same as the reference, and 1 means totally different. 
During each MOS evaluation, the speaker similarity or style similarity is the only point that testers need to focus on.
%During speaker similarity and style similarity MOS evaluation, we asked the testers not to consider the content to which they are listening and only pay attention to speaker similarity or style similarity.
%MOS of naturalness is not considered in this experiment, as the focus of this paper is the disentanglement of a speaker's timbre and style, although the quality of the synthesized speech is state-of-the-art for cross-speaker style transfer.
% As the synthesized speech with any combination of speaker and style is natural enough, and the focus of this paper is the disentanglement of speaker and timbre, naturalness MOS evaluation is not conducted currently.
% a 5 points MOS evaluation is used where 1 means bad, 2 means poor, 3 means fair, 4 means good and 5 means excellent.

\begin{table}[t]
  % \caption{The MOS results of speaker similarity and style similarity with a confidence interval of 95\%, three open-sourced speakers are selected as target speakers and speech of three different styles from the internal corpus are synthesized and evaluated.}
    \caption{The 5 points MOS results with a confidence interval of 95\%. 5 means the style or the timbre is exactly the same as the reference, and 1 means totally different.}
%   \vspace{3pt}
  \label{tab:mos_results}
  \centering
  \begin{tabular}{l|ccc|ccc}
  \hline
  \multirow{2}{*}{\centering{\textbf{Speaker}}} &\multicolumn{3}{c}{Speaker Similarity}\vline & \multicolumn{3}{c}{Style Similarity}      \\
  
    \cline{2-7}          &C1           &F1             &M1             &C1          &F1             &M1             \\
    \hline
    CSMSC            &4.50$\pm$0.05   &4.47$\pm$0.06  &4.49$\pm$0.04  &4.31$\pm$0.03  &4.62$\pm$0.03  &4.11$\pm$0.06  \\
    \hline
    Originbeat-S1    &4.36$\pm$0.08    &4.66$\pm$0.02       &4.29$\pm$0.06       &4.13$\pm$0.03          &4.18$\pm$0.05       &4.08$\pm$0.08       \\
    \hline
    Originbeat-S2    &4.36$\pm$0.07    &4.59$\pm$0.04        &4.23$\pm$0.07       &4.23$\pm$0.04          &4.30$\pm$0.03      &4.10$\pm$0.08        \\
  \hline
  \end{tabular}
  % \vspace{-3mm}
\end{table}

MOS results of speaker similarity and style similarity are shown in Table~\ref{tab:mos_results}, we could see that for a given target speaker, the speaker similarity of speeches in different style are very high and consistent, and the style similarity is also very high with the target style, which indicates that the proposed model achieves excellent disentanglement effect for speaker's timbre and style.
%MOS results of speaker similarity and style similarity are shown in Table~\ref{tab:mos_results}, from this table we could see that for a given target speaker, the speaker similarity of speeches in different style are very high and consistent, which showed the effectiveness of the speaker timbre information retention; and the style similarity is also very high with the target style, which indicates that the proposed model in this paper achieves excellent disentanglement effect for speaker's timbre and style.

\subsection{Ablation Study Of Utterance Level Feature Normalization}

To verify the effectiveness of UttNorm, another female corpus(112 sentences) crawled from a podcast, even with some background noise is chosen to demonstrate the impact. 
%This noisy corpus is crawled from a podcast with high emotion and style variation, and even with some background noise. There are total 112 sentences in this noisy corpus. 
%Experiments with speaker-level normalization and utterance-level normalization are conducted to evaluate the impact of different feature normalization methods.

In our experiments, due to the large variety of this noisy corpus from the oral podcast, a speaker similarity MOS evaluation and a preference evaluation are conducted to evaluate the performance of different normalization methods. We use speaker F1 as the target speaker and the style from this noisy corpus as the target style, then 20 sentences in the podcast domain are synthesized and listened to by 15 testers. 
%Speaker similarity and style similarity MOS evaluations are conducted to evaluate the performance of different normalization method, we use speaker F1 as target speaker and the style from this noisy corpus as the target style, then 20(xxx) sentences in podcast domain are synthesized and listened by 15 (xxx) testers. There are five points in the MOS evaluation, ranging from 1 to 5.

% From table~\ref{tab:mos_ablation_utt_norm} we could see that when utterance level feature normalization is used, the style similarity MOS increases by XX\%, which showed that utterance level feature normalization could facilitate the decoupling effect of timbre and style for data with high variance, especially found data, noisy data and podcast data.
% The speaker similarity is nearly the same for the two feature normalization methods, this is because the speaker's timbre is learned by the clean F1 corpus, and the proposed model only learns style from this noisy corpus, the introduction of noisy corpus does not affect the timbre of F1.

According to the MOS results for speaker similarity, when UttNorm is used, the speaker similarity increases from 3.82$\pm$0.05(SpkNorm) to 3.91$\pm$0.06(UttNorm), which showed that UttNorm could facilitate the decoupling effect of timbre and style for data with high variance, especially found data, noisy data or podcast data.
% The speaker similarity is nearly the same for the two feature normalization methods, this is because the speaker's timbre is learned by the clean F1 corpus, and the proposed model only learns style from this noisy corpus, the introduction of noisy corpus does not affect the timbre of F1.

From the AB preference evaluation results, the proportions of preference are 0.17(SpkNorm), 0.35(No Preference), 0.48(UttNorm), and the \textit{p}-value is less than 0.001. It could be found that the listeners strongly prefer the results from UttNorm, which proves the effectiveness of utterance normalization.

\subsection{Demonstration Of The Proposed Model}

\begin{figure}
\vspace{-3mm}
\centering
\includegraphics[width=\textwidth]{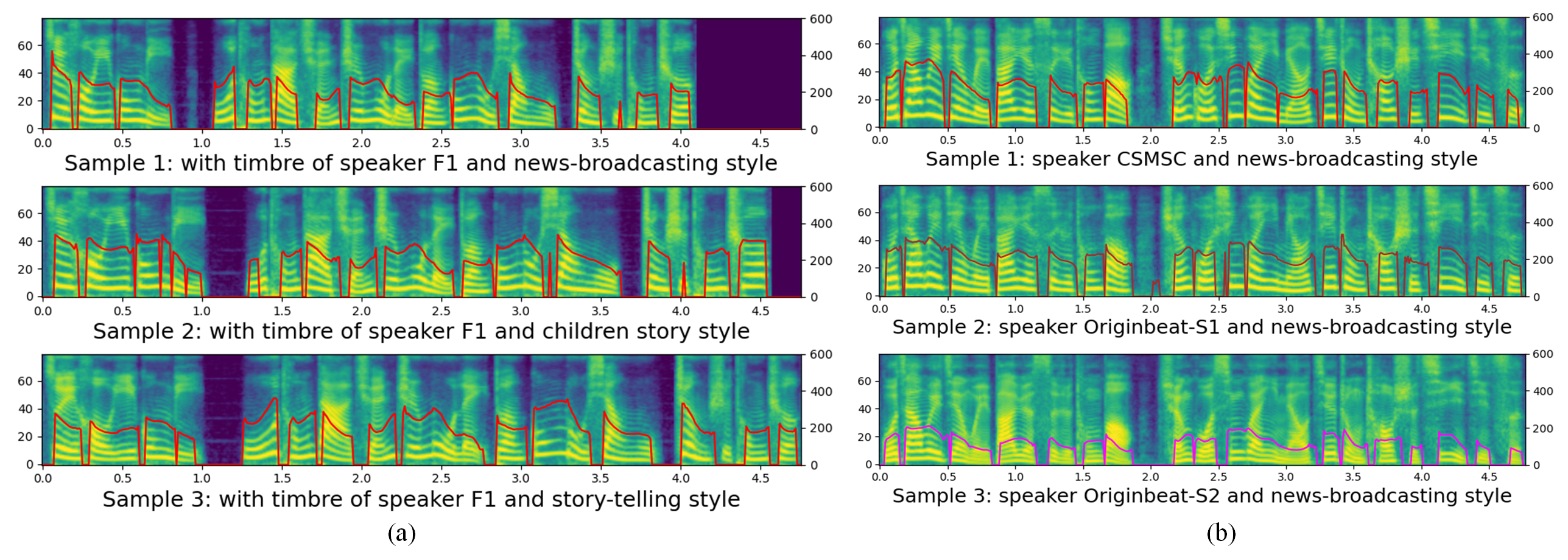}
\caption{A mel-spectrogram illustration of the proposed multi-speaker multi-style TTS model. (a) Speaker F1 is selected as the target speaker, the target style is the original news-broadcasting style, children story style, and story-telling style. (b) News-broadcasting style is selected as target style, the speakers are CSMSC, Originbeat-S1, and Originbeat-S2 respectively, from top to bottom.}
\vspace{-5mm}
\label{different_style}
\end{figure}

Synthesized speeches with a given speaker and different styles are shown in Fig.~\ref{different_style} (a), and speeches for a given style and different speakers are shown in Fig.~\ref{different_style} (b). 
The fundamental frequency is consistent for the three different styles in Fig.~\ref{different_style} (a), and all the speakers' pitch trajectories follow basically the same curve but with different fundamental frequency in Fig.~\ref{different_style} (b), this explains the decoupling effectiveness to some extent. 
% which showed that the style is almost the same in these 3 different speeches. 
% this explains the decoupling effectiveness to some extent. 

% % Mel-spectrograms of speeches for a given style and different speakers are shown in Fig.~\ref{different_style}(b).
% All the speakers' pitch trajectories follow basically the same curve which showed that the style is almost the same in these 3 different speeches. 
% The different F0 values in each speaker's speech showed that the proposed model indeed synthesized speech with a given timbre and keep the style consistent.

% \begin{figure}
% \vspace{-3mm}
% \centering
% \includegraphics[width=0.8\textwidth]{figs/same_speaker_different_style.png}
% \caption{An illustration of mel-spectrogram with different styles. In this figure, speaker F1 is selected as the target speaker, the target style is the speaker's original news-broadcasting style, children story style, and story-telling story, from top to bottom respectively.}
% \vspace{-5mm}
% \label{different_style}
% \end{figure}

% \begin{figure}[t]
% \centering
% \includegraphics[width=0.8\textwidth]{figs/different_speaker_same_style.png}
% \caption{The mel-spectrogram demonstration of synthesized speech with different speakers and identical style. From top to bottom, the speakers are CSMSC, Originbeat-S1, and Originbeat-S2 respectively.}
% \vspace{-5mm}
% \label{different_speaker}
% \end{figure}

\subsection{Style Transition Illustration}

% After the disentanglement of style and timbre, the proposed speech synthesis system could achieve gradual style transition from source speaker's style to target speaker's style, by adjusting the combination weights of both the source speaker's and target speaker's style embedding, while keeping the source speaker's timbre unchanged.

As both a speaker ID and a style ID should be sent to the network to synthesize the target speaker's speech with a given style, we could also use the speaker ID to generate a style embedding that represents the source style, and the embedding from the style ID represents the target style. 
By combining the style embeddings from source and target with different weights, we could generate speech with different target style intensities.

\begin{figure}[t]
\centering
\includegraphics[width=0.9\textwidth]{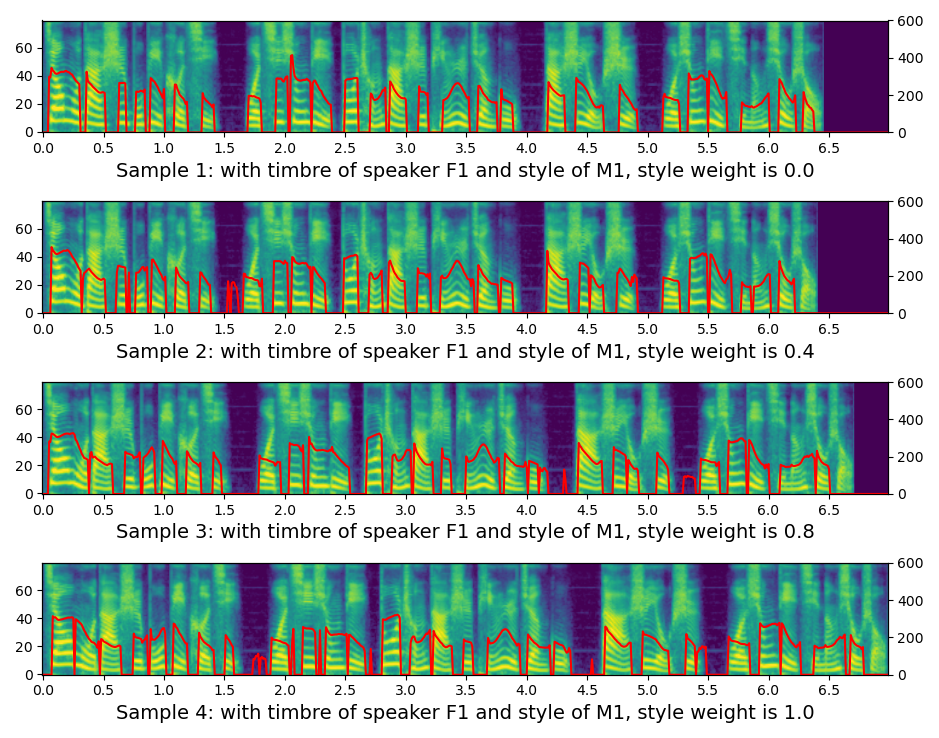}
\caption{A style transition illustration. Speaker F1 is selected as the target speaker and style M1 is selected as the target style, this figure shows the gradual style transition from style F1 to style M1 from top to bottom, with different target style weights.}
\vspace{-5mm}
\label{style_transition}
\end{figure}

To demonstrate the continuity of the learned style embedding representation, a style transition example is given in Fig.~\ref{style_transition}. 
% speaker F1 is used as the target speaker, and the style gradually transits from source style (style F1) to the target M1 style from top to bottom, the text in each synthesized speech is identical. 
From the pitch trajectory in each synthesized speech we could find that the fundamental frequency is identical for different target style weight, indicating that the proposed model keep the timbre unchanged when synthesizing different style speech and achieves outstanding timbre and style disentanglement effect.

\section{Conclusions}

A simple but effective speaker's timbre and style disentanglement network is proposed in this paper, which eliminates the reliance on a single-speaker multi-style corpus. 
The proposed network learns a style-dependent variance adaptor and a speaker-dependent mel-spectrogram prediction decoder. 
% , by carefully designing the network structure.
The style-related features are predicted by the variance adaptor with the guidance of style embedding, while the timbre is learned by the mel-spectrogram decoder with the control of speaker embedding.
% the speaker embedding and style embedding are carefully incorporated into the network to learn decoupled style and timbre representation. 
Utterance level feature normalization is proposed to prevent speaker information leakage from the style feature.
Experimental results showed that the proposed model achieves good timbre and style disentanglement effect, for a given speaker the proposed model could synthesize speech with any style seen during training, even when the target style corpus only contains a few hundred training utterances. 
Furthermore, the proposed model learns a continuous style representation, which could generate speech that gradually transits from source style to target style.
% When noisy data is used to learn a style, the proposed utterance level feature normalization greatly improved the speaker similarity and style similarity for the synthesized speech.

\newpage
\vfill\pagebreak


% \scriptsize
% \bibliographystyle{unsrt}
\bibliographystyle{IEEEbib2}
% \footnotesize
% \tiny
\bibliography{ref}
\end{document}